\documentclass{elsart}
\journal{New Astronomy}

\usepackage{natbib}

\usepackage{graphicx}
\usepackage{epsfig}
\usepackage{amssymb}

\begin{document}

\begin{frontmatter}

 \title{Star Forming Galaxies at $z\approx 6$ and Reionization}

\author{Andrew Bunker$^1$, Elizabeth Stanway$^2$, Richard Ellis$^3$,}
\author{Richard McMahon$^4$, Laurence Eyles$^1$ and Mark Lacy$^3$}
\address{$^1$University of Exeter, School of Physics, Stocker Road, Exeter, 
EX4\,4QL, U.K.\\
$^2$Astronomy Department, University of Wisconsin-Madison, 475 N.\ Charter Street, Madison, WI~53706, U.S.A.\\
$^3$California Institute of Technology, Pasadena, CA 91125, U.S.A.\\
$^4$Institute of Astronomy, Madingley Road, Cambridge CB3 0HA, 
U.K.\\
E-mail: bunker@astro.ex.ac.uk, stanway@astro.wisc.edu, rse@astro.caltech.edu, 
rgm@ast.cam.ac.uk, eyles@astro.ex.ac.uk, mlacy@ipac.caltech.edu}

\begin{abstract}
  We have determined the abundance of $i'$-band drop-outs in the {\em
  HST}/ACS GOODS surveys and the Hubble Ultra Deep Field (UDF).  The
  majority of these sources are likely to be $z\approx 6$ galaxies
  whose flux decrement between the F775W $i'$-band and F850LP
  $z'$-band arises from Lyman-alpha absorption. We have shown with
  Keck/DEIMOS and Gemini/GMOS spectroscopy that this technique does
  indeed select high redshift galaxies, and we discovered
  Lyman-$\alpha$ emission in the expected redshift range for about a
  third of the galaxies with $z'_{AB}<25.6$ in the 150\,arcmin$^2$ of
  the GOODS-South field.  
The $i$-drop number counts in the GOODS-North field are consistent, so
  cosmic variance is possibly not the dominant uncertainty. The
  increased depth of UDF enables us to reach a $\sim 10\,\sigma$
  limiting magnitude of $z'_{AB}=28.5$ (equivalent to
  $1.5\,h^{-2}_{70}\,M_{\odot}\,{\rm yr}^{-1}$ at $z=6.1$, or
  $0.1\,L^{*}_{UV}$ for the $z\approx 3$ $U$-drop population).
	The star formation rate at $z\approx 6$ was approximately
  $\times 6$ {\em less} than at $z\approx 3$. This declining comoving
  star formation rate ($0.005\,h_{70}\,M_{\odot}\,{\rm yr}^{-1}\,{\rm
  Mpc}^{-3}$ at $z\approx 6$ at $L_{UV}>0.1\,L^{*}$ for a Salpeter
  IMF) poses an interesting challenge for models which suggest that
  $L_{UV}>0.1\,L^{*}$ star forming galaxies at $z\simeq 6$ reionized
  the universe. The short-fall in ionizing photons might be alleviated
  by galaxies fainter than our limit, or a radically different IMF.
  Alternatively, the bulk of reionization might have occurred at
  $z\gg6$. We have recently discovered evidence of an early epoch of
  star formation in some of the $i'$-drops at
  $z\approx 6$.  {\em Spitzer} images with IRAC at $3.6-4.5\,\mu$m
  show evidence of the age-sensitive Balmer/4000\,\AA , dominated by
  stars older than 100\,Myr (and most probably 400\,Myr old). This
  pushes the formation epoch for these galaxies to $z_{\rm
  form}=7.5-13.5$. There are at least some galaxies already assembled
  with stellar masses $\approx 3\times 10^{10}\,M_{\odot}$ (equivalent
  to $0.2\,M^{*}$ today) within the first billion years. The early
  formation of such systems may have played a key role in reionizing
  the Universe at $z\sim 10$.

\end{abstract}

%

\end{frontmatter}

\section{Introduction}

There has been enormous progress over the past decade in discovering
galaxies and QSOs at increasingly high redshifts. We are now probing
far enough back in time that the Universe at these early epochs was
fundamentally different from its predominantly ionized state today.
Observations of $z>6.2$ QSOs (Becker et al.\ 2001, Fan et al.\ 2002)
show near-complete absorption of flux at wavelengths short-ward of
Lyman-$\alpha$ (Gunn \& Peterson 1965), indicating that the Universe
is optically thick to this line, and that the neutral fraction of
hydrogen is much greater than at lower redshifts. Initial cosmic
microwave background results from the {\em WMAP} satellite indicated
that the Universe was completely neutral at redshifts of $z\sim 10$
(Kogut et al.\ 2003).  There is an ongoing debate as to what reionized
the Universe at $z>6$: is it AGN or ionizing photons (produced in hot,
massive, short-lived stars) escaping from star forming galaxies? Along
with the escape fraction for these ionizing photons, it is crucial to
know the global star formation rate at this epoch.

Here we detail our work in identifying star-forming galaxies within
the first billion years of the Big Bang through deep imaging with the
{\em Hubble Space telescope (HST)}.  Our selection method is based on
the Lyman break technique, and we have shown this is effective in
isolating $z\approx 6$ by obtaining spectroscopic confirmation with
the 10-m Keck telescopes. Our analysis of the {\em Hubble Ultra
  Deep Field} (the deepest images ever obtained) enables us to explore
the faint end of the luminosity function (the bulk of galaxies with
low star formation rates). Our discovery of this $i'$-drop galaxy
population is used to infer the global star formation rate density at
this epoch ($z\approx 6$).  We consider the implications for
reionization of the UV flux from these galaxies. We also use infrared
data from the {\em Spitzer Space Telescope} to determine the spectral
energy distributions (SEDs) of our $i'$-drops, and constrain
the previous star formation histories of these sources.

Throughout we adopt the
standard ``concordance'' cosmology of $\Omega_M=0.3$,
$\Omega_{\Lambda}=0.7$, and use $h_{70}=H_0/70\,{\rm
km\,s^{-1}\,Mpc^{-1}}$. All magnitudes are on the $AB$ system.

\section{The Lyman-Break Technique at $z\approx 6$}

\begin{figure}
\resizebox{0.49\textwidth}{!}{\includegraphics{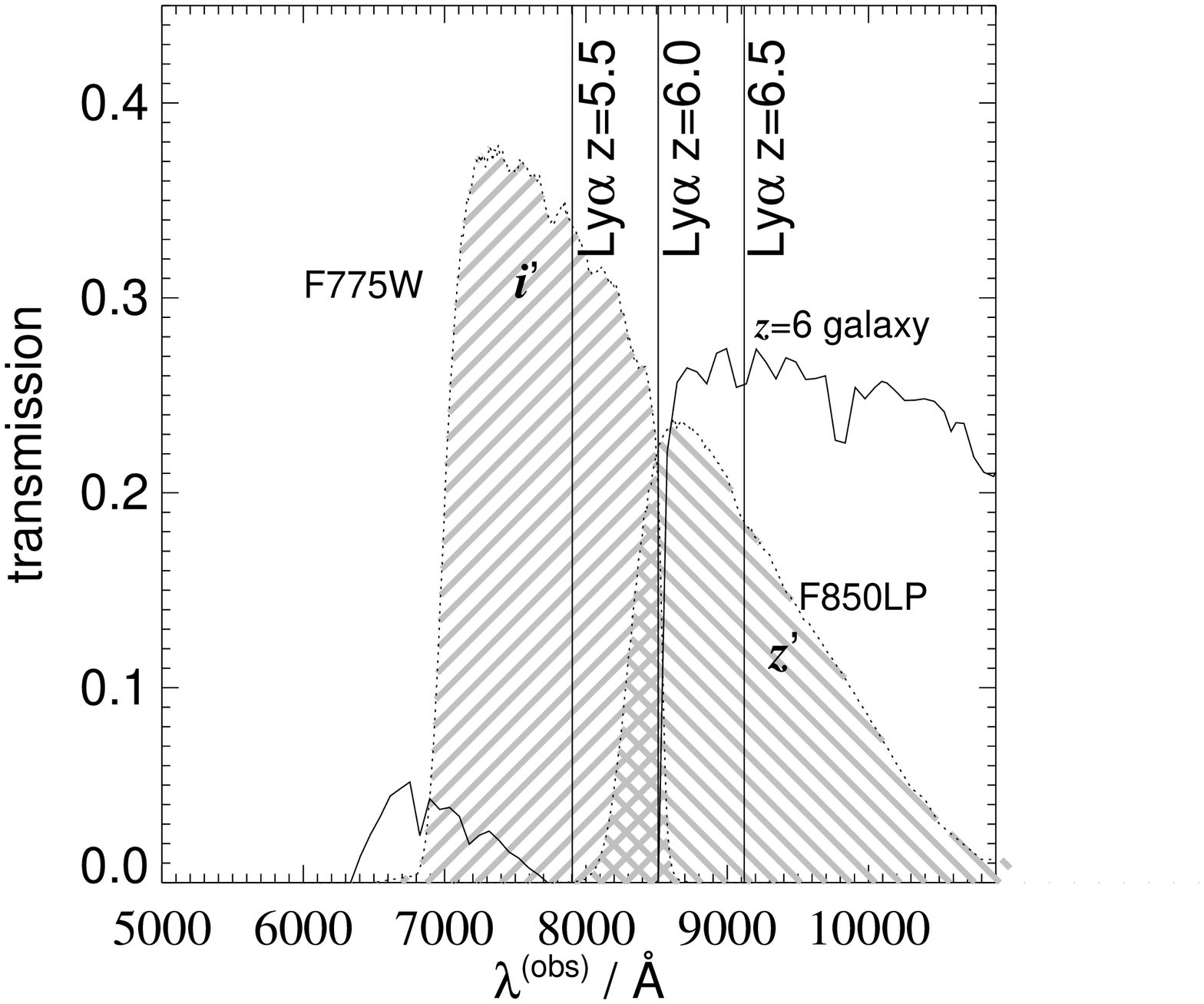}}
\resizebox{0.49\textwidth}{!}{\includegraphics{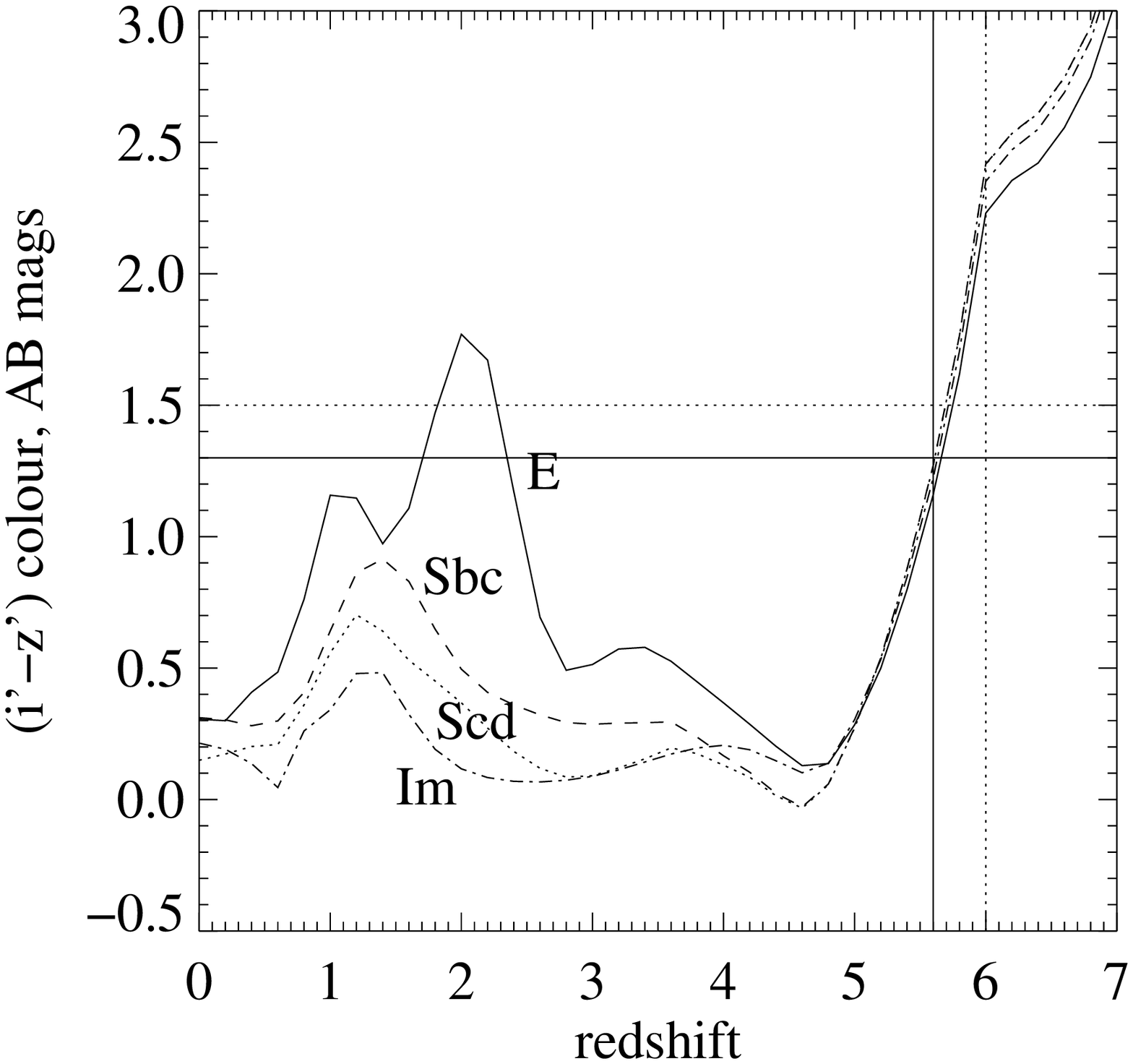}}
\caption{{\bf Left:} The ACS-$i'$ and -$z'$ bandpasses overplotted on
the spectrum of a generic $z=6$ galaxy (solid line), illustrating the
utility of our two-filter technique for locating $z\approx 6$
sources. {\bf Right:} Model colour-redshift tracks for galaxies with
non-evolving stellar populations (from the template spectra of Coleman,
Wu \& Weedman 1980, ApJS 43, 393 ). The contaminating `hump' in the
$(i'-z')$ colour at $z\approx 1-2$ arises when the Balmer break and/or
the 4000\,\AA\ break redshifts beyond the $i'$-filter.}
\label{fig:filters}
\label{fig:tracks}
\end{figure}

\begin{figure}
\resizebox{0.49\textwidth}{!}{\includegraphics*[65,25][475,465]{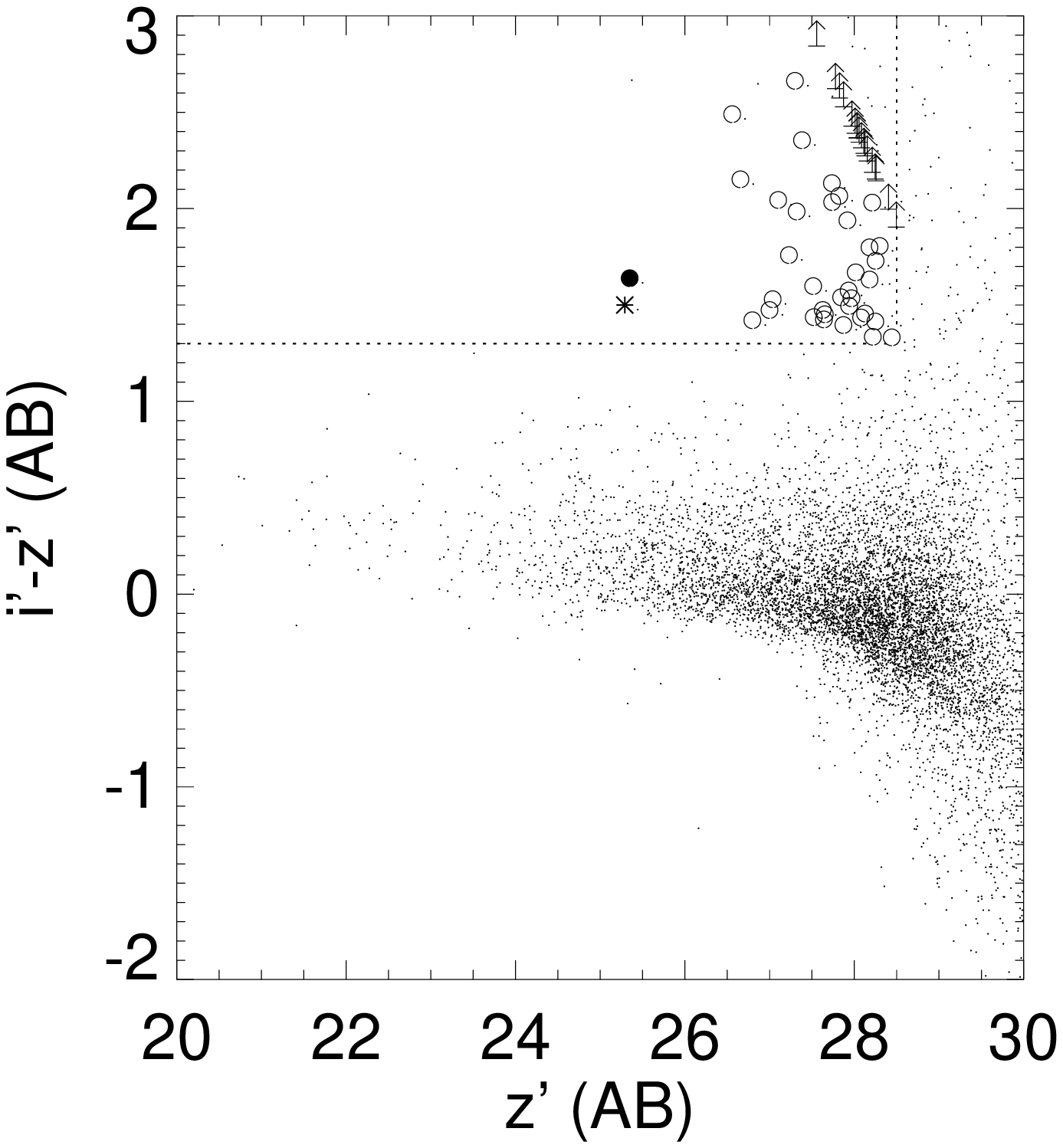}}
\resizebox{0.49\textwidth}{!}{\includegraphics{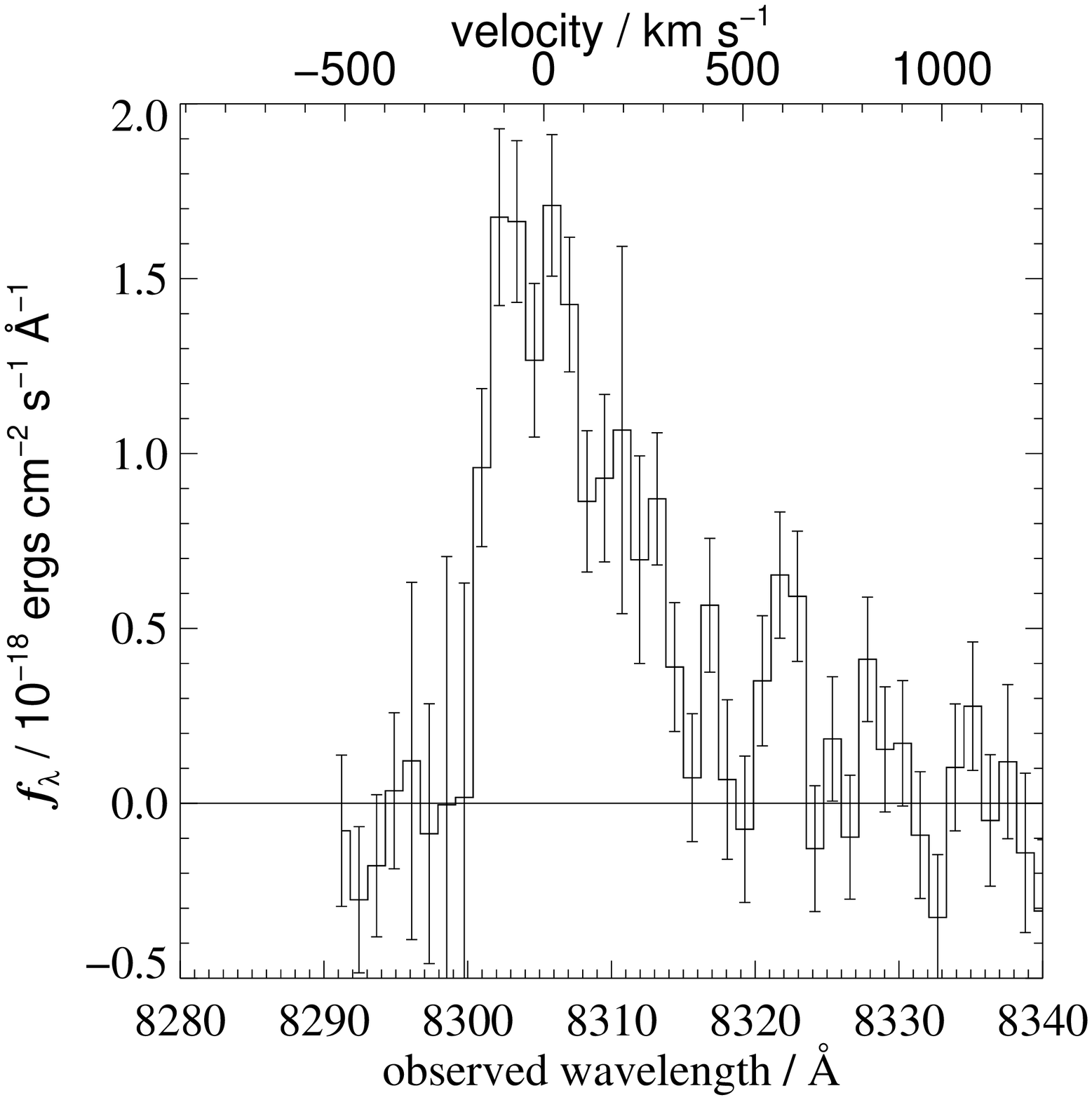}}
\caption{{\bf Left:} Colour-magnitude diagram for the UDF data with the
limit $z'_{AB}<28.5$ and $(i'-z')_{AB}=1.3$ colour cut shown (dashed
lines). Such a catalogue could be contaminated by cool stars, EROs and
wrongly identified extended objects and diffraction spikes but
nonetheless provides a secure upper limit to the abundance of z$\approx
6$ star forming galaxies.  Circles and arrows (lower limits) indicate
our $i'$-drop candidate $z\approx 6$ galaxies. The asterisk is the only
unresolved $i'$-drop in our UDF sample, a probable star. The solid
circle is the brightest $i'$-drop in the UDF, the
spectroscopically-confirmed galaxy SBM03\#1 at $z=5.83$, with our
discovery spectrum from Keck/DEIMOS shown ({\bf right}, Stanway et al.\
2004b.}
\label{fig:izzcolmag}
\end{figure}

In order to select $z\approx 6$ galaxies, we use the ``Lyman break
technique'' pioneered at $z\sim 3$ using ground-based telescopes by
Steidel and co-workers (Steidel, Pettini \& Hamilton 1995; Steidel et
al.\ 1996) and using {\em HST} by Madau et al.\ (1996).  At $z\approx
6$, we can efficiently use only two filters, above ($z'$-band 9000\AA)
and below ($i'$-band 8000\AA ) the continuum break at Lyman-$\alpha$
line ($\rm\lambda =(1+z)\times1216$\,\AA ) since the integrated
optical depth of the Lyman-$\alpha$ forest is $\gg 1$ (see
Figure~\ref{fig:filters}) and there is essentially no flux at shorter
wavelengths.  The key issue is to work at a sufficiently-high
signal-to-noise ratio that Lyman break ``drop-out'' galaxies can be
safely identified through detection in a single redder band (i.e.,
$z'$-band).  This approach has been demonstrated to be effective by
the SDSS collaboration in the detection of $z\approx 6$ QSOs using the
$i'$- and $z'$-bands alone (Fan et al.\ 2002), using wide-area but
shallow ground-based imaging (sensitive to bright but rare quasars).
Our group has analysed deep imaging from the Advanced Camera for
Surveys (ACS) on {\em HST} with the same sharp-sided SDSS F775W ($i'$)
and F850LP ($z'$) filters to locate the``$i$-drops'' which are
candidate $z\approx 6$ galaxies (fainter but more numerous than the
QSOs). In Figures~\ref{fig:tracks}\,\&\,\ref{fig:izzcolmag} we
illustrate how a colour cut of $(i'-z')_{AB}>1.3$ can be effective in
selecting sources with $z>5.6$.

We have analysed the ACS images from the {\em HST} Treasury ``Great
Observatory Origins Deep Survey'' (GOODS; Giavalisco \& Dickinson
2003) to discover this $i'$-drop population of $z\approx 6$ galaxies.
In Stanway, Bunker \& McMahon (2003) we searched 150\,arcmin$^{2}$ of
the GOODS-South field (the {\em Chandra} Deep Field-South) to select
$(i'-z')_{AB}>1.5$ objects, of which six were probable $z>5.7$
galaxies brighter than $z_{AB}<25.6$.  Subsequent papers by the GOODS
team (Dickinson {\em et al.\ }2004; Giavalisco {\em et al.\ }2004)
reproduced many of these $i'$-drop candidates.  To address potential
cosmic variance issues, we performed a similar analysis in the
GOODS-North field, which yielded a consistent estimate of the surface
density of $z\simeq 6$ star forming sources (Stanway et al.\ 2004b).

\subsection{Spectroscopic Confirmation of $z\approx 6$ Galaxies}

The effectiveness of the Lyman-break technique has been demonstrated
at lower redshifts through spectroscopic confirmation of hundreds of
$U$-band dropouts at $z\approx 3$ and $B$-band drop-outs at $z\approx
4$.  However, it is important to demonstrate that the $i'$-drop
selection is similarly isolating galaxies at $z\approx 6$, if we are
to use the surface density of $i'$-drops to draw global inferences
about galaxies within the first billion years. A known possible
contaminant is the Extremely Red Object (ERO) population of evolved
galaxies at $z\approx 1-2$ which can produce large $(i'-z')$ colours:
deep near-infrared imaging (e.g., from the {\em HST}/NICMOS survey of
the UDF) should identify EROs, and we consider this possible
contamination Stanway, McMahon \& Bunker (2005). Low-mass Galactic
stars (M/L/T dwarfs) are another interloper population.

We have obtained deep spectroscopy with Keck/DEIMOS of several of the
brightest $i'$-drops identified by Stanway, Bunker \& McMahon (2003)
in GOODS-South; in the case of SBM03\#3 ($z'=25.6$) and , an $i'$-drop
in the Chandra Deep Field South photometrically-selected from HST/ACS
images to lie at $z\approx 6$.

For two of these objects, we discovered single emission lines (Fig.~3)
at 8245\,\AA\ \& 8305\,\AA , consistent with Lyman-$\alpha$ emission
from galaxies at $z=5.78$ (SBM03\#3, Bunker et al.\ 2003) and $z=5.83$
(SBM03\#1, Stanway et al.\ 2004a). The spectrally-resolved profiles of
the emission lines are asymmetric (as high-$z$ Lyman-$\alpha$ tends to
be) with a P-Cygni-like profile and a sharp cut-off on the blue wing
(Fig.~2).  The line fluxes are $\approx 2\times 10^{-17}\,{\rm
ergs\,cm}^{-1}\,{\rm s}^{-1}$. The equivalent widths are $W_{\rm
rest}=20-30$\,\AA\ using the $z'$-band photometry from HST/ACS, which
is within the range seen in high-$z$ star-forming galaxies at $z\sim
3-4$. The velocity widths of the Lyman-$\alpha$ are comparatively
narrow ($v_{\rm FWHM}=300\,{\rm km\,s}^{-1}$) and we do not detect the
high-ionization line N{\scriptsize~V}\,$\lambda$\,12140\,\AA , which
strongly support the view that this line emission is powered by star
formation rather than an AGN. Our spectroscopic redshifts for these
objects confirms the validity of the $i'$-drop selection technique to
select star-forming galaxies at $z\approx 6$.  Subsequently, a
programme of spectroscopy with GMOS ``nod\,\&\,shuffle'' on Gemini
(GLARE, Stanway et al.\ 2004b) has confirmed more galaxies in
GOODS-South. Slitless spectroscopy of the UDF with {\em HST}/ACS
(GRAPES, Malhotra et al.\ 2005) has also confirmed the spectral breaks
of the $i'$-drop galaxies, although the spectral resolution of
$R\approx 70$ is insufficient to see the Lyman-$\alpha$ emission
lines.

\begin{figure}
\resizebox{0.98\textwidth}{!}{\includegraphics{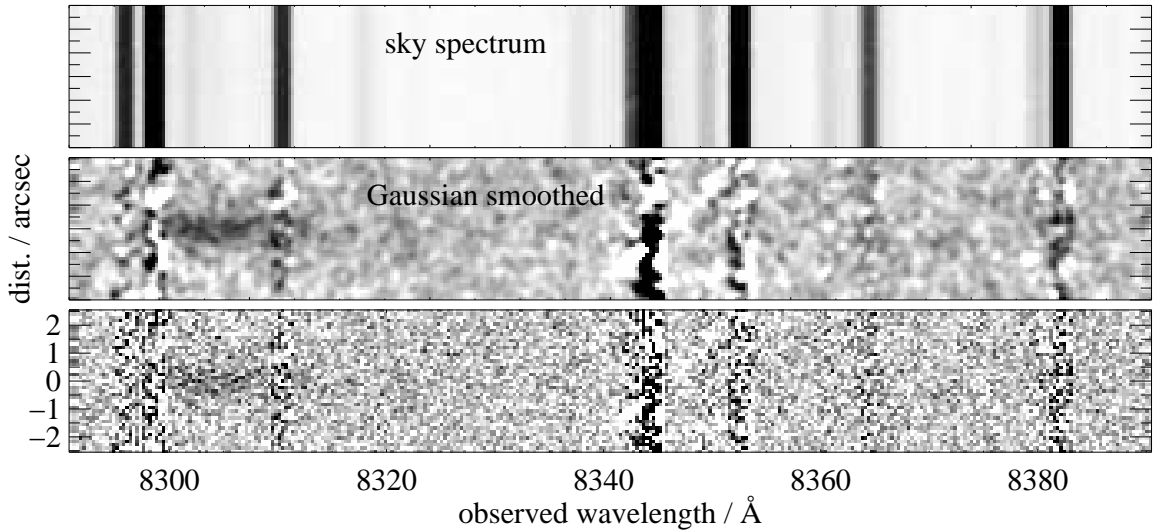}}
\caption{Our 2D Keck/DEIMOS spectrum confirming the brightest
$i'$-drop in the UDF is indeed a $z\approx 6$ galaxy: prominent
Lyman-$\alpha$ emission is seen at $z=5.83$ (from Stanway et al.\
2004b). There has been independent spectroscopic confirmation by
Dickinson et al.\ (2004).}
\label{fig:2Dspec}
\end{figure}
\begin{figure}
\resizebox{0.49\columnwidth}{!}{\includegraphics{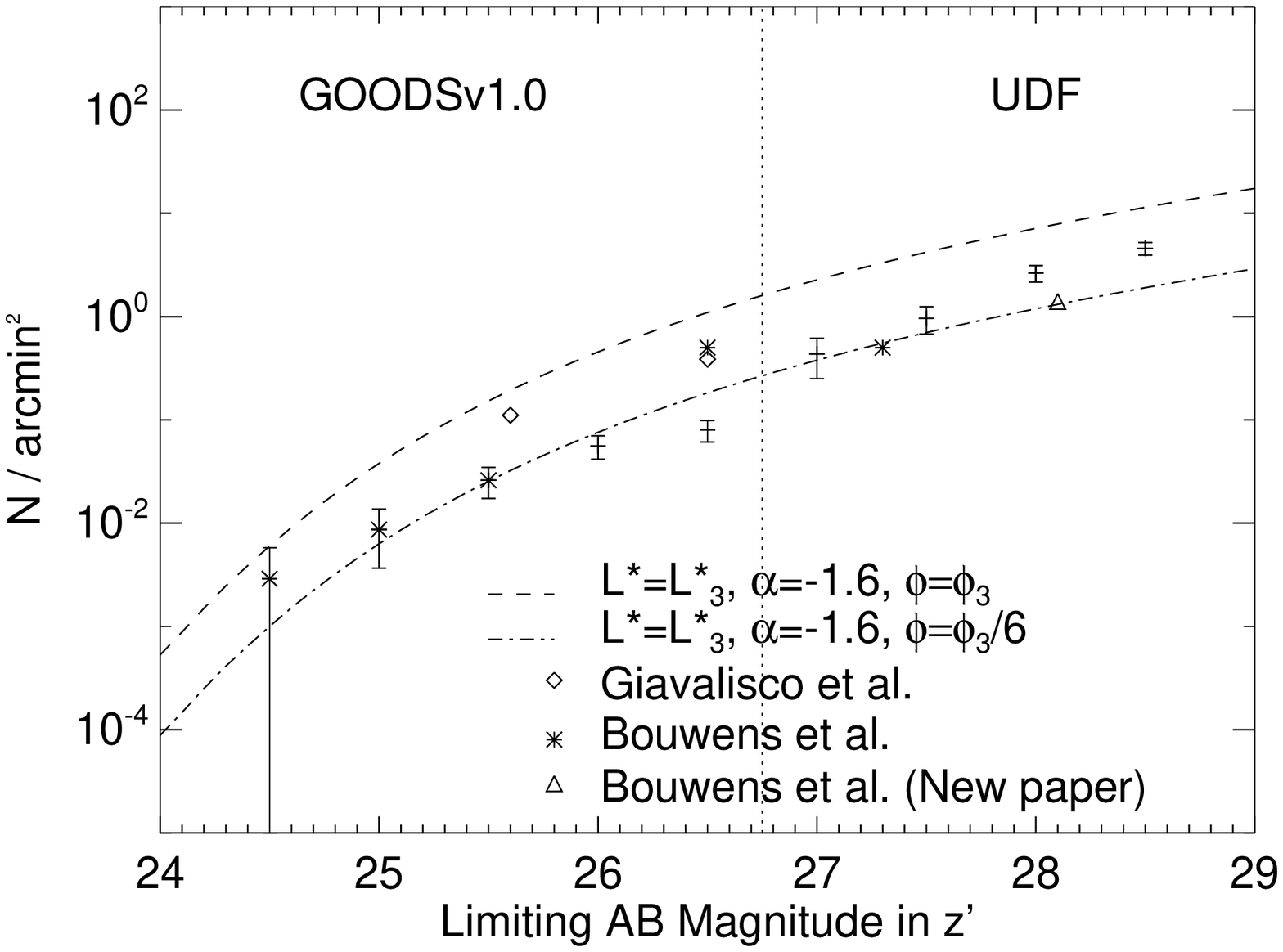}}
\resizebox{0.49\textwidth}{!}{\includegraphics{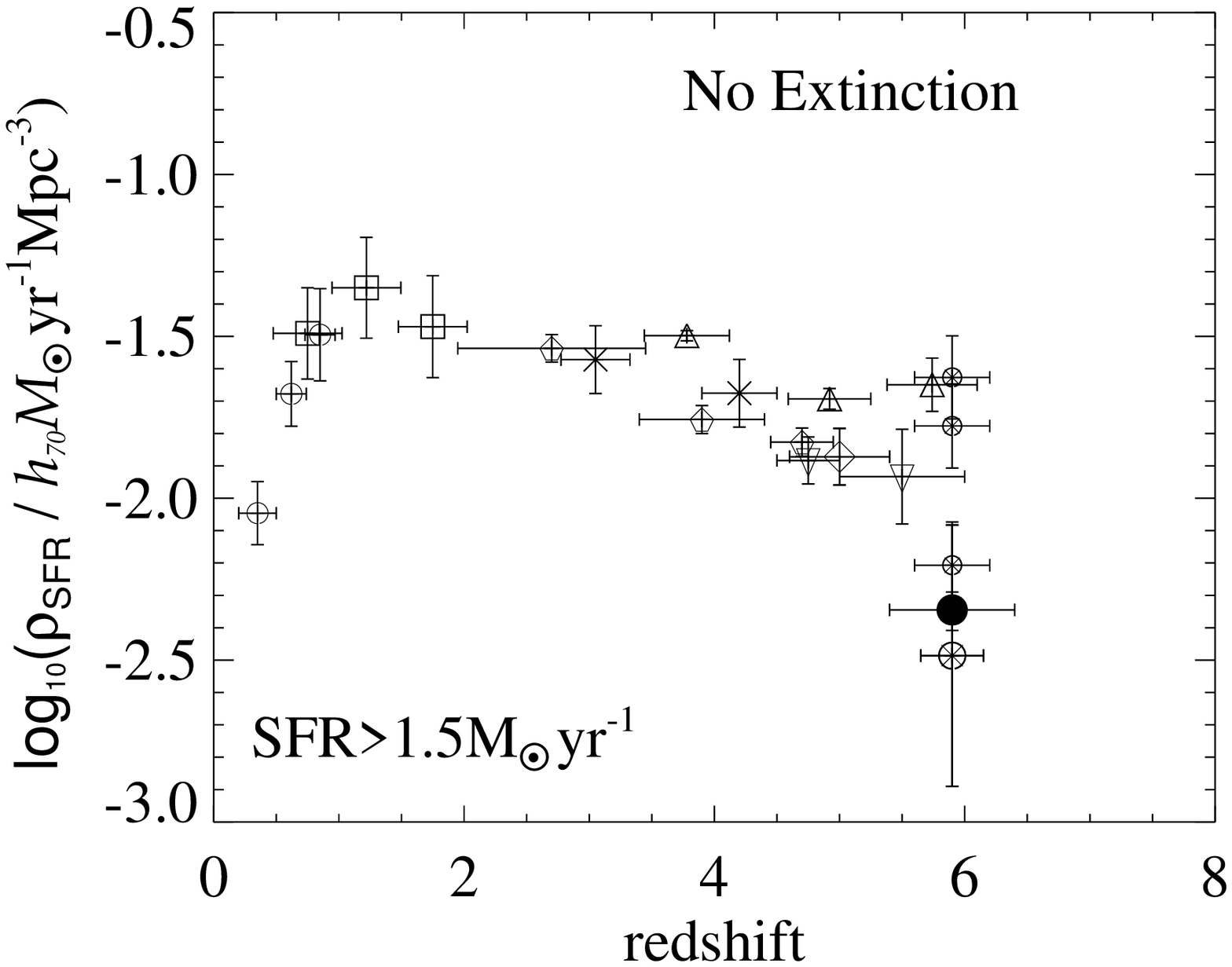}}
\caption{{\bf Left:} Cumulative source counts per arcmin$^{2}$ of
$i'$-dropouts as a function of $z'$-band magnitude. The new UDF data
(over an area of 11\,arcmin$^{2}$ for $28.5\ge z'_{AB}\ge 27.0$) is compared
with our $z'_{AB}\le 25.6$ single epoch GOODSv0.5 ACS/WFC imaging over
300\,arcmin$^{2}$ (Stanway, Bunker \& McMahon 2003; Stanway et al.\
2004), and combined 5 epoch GOODSv1.0 images to $z'_{AB}<27.0$
(Stanway 2004, PhD.\ thesis, Cambridge University). Also plotted are
the surface densities of $i'$-drops from Giavalisco et al.\ (2004)
Bouwens et al.\ (2003, 2004)
\newline
{\bf Right:} An updated version of the `Madau-Lilly' diagram (Madau et
al.\ 1996; Lilly et al.\ 1996) illustrating the evolution of the
comoving volume-averaged star formation rate.  Our work from the UDF
data is plotted a solid symbol. Other determinations have been
recalculated for our cosmology and limiting UV luminosity of
$1.5\,h_{70}^{-2}\,M_{\odot}\,{\rm yr}^{-1}$ at $z=6.1$ (equivalent to
$0.1\,L^*_3$ at $z\approx 3$ from Steidel et al.\ 1999), assuming a
slope of $\alpha=-1.6$ for $z>2$ and $\alpha=-1.3$ for $z<2$. Data
from the CFRS survey of Lilly et al.\ (1996) are shown as open
circles; data from Connolly et al.\ (1997) are squares; and the Lyman
break galaxy work of Steidel et al.\ (1999) is plotted as crosses, of
Fontana et al.\ (2002) as inverted triangles and that by Iwata et al.\
(2003) as an open diamond. Pentagons are from Bouwens, Broadhurst \&
Illingworth (2003), upright triangles are the GOODS $i'$-drop results
from Giavalisco et al.\ (2004).  The three ACS estimates of Bouwens et
al.\ (2003) are shown by small crossed circles and indicate three
different completeness corrections for one sample of objects -- the
larger symbol is the recent re-determination using a new catalogue by
this group from a deeper dataset (the UDF flanking fields -- Bouwens
et al.\ 2004).}
\label{fig:numcounts}
\label{fig:madau_plot}
\end{figure}

\section{The star formation rate density at $z\approx 6$}
\label{sec:pol}

We base our measurement of the star formation rate on the intensity of
the rest-frame UV continuum emission, redshifted into the $z'$-band at
$z\approx 6$; this is dominated by the light from the most massive
stars ($M>10\,M_{\odot}$ which are the hottest and bluest stars),
which are the shortest lived, and hence a tracer of the instantaneous
star formation rate. To go from the number of massive stars formed to
the total star formation rate requires an assumption about the stellar
initial mass function (IMF). In the absence of dust obscuration, the
relation between the flux density in the rest-UV around $\approx
1500$\,\AA\ and the star formation rate (${\rm SFR}$ in
$M_{\odot}\,{\rm yr}^{-1}$) is given by $L_{\rm UV}=8\times 10^{27}
{\rm SFR}\,{\rm ergs\,s^{-1}\,Hz^{-1}}$ from Madau, Pozzetti \&
Dickinson (1998) for a Salpeter (1955) IMF with
$0.1\,M_{\odot}<M^{*}<125\,M_{\odot}$. Our relatively bright magnitude
cut of $z'_{AB}<25.6$ for the GOODSv0.5 data (Stanway, Bunker \&
McMahon 2003; Stanway et al.\ 2004a) corresponds to an unobscured star
formation rate of $15\,h^{-2}_{70}\,M_{\odot}\,{\rm yr}^{-1}$ at
$z=5.9$ (the luminosity-weighted average redshift), equivalent to
$L^{*}_{UV}$ for the $U$-band dropout population at $z\approx 3$.

We recover the $z\approx 6$ rest-frame UV luminosity function from the
observed number counts of $i'$-drops to faint magnitudes in the UDF.
Although our colour cut selects galaxies with redshifts in the range
$5.6<z<7.0$, an increasing fraction of the $z'$-band flux is
attenuated by the redshifted Lyman-$\alpha$ forest. At higher
redshifts we probe increasingly shortward of $\lambda_{\rm
rest}=1500$\,\AA\ (where the luminosity function is calculated) so the
$k$-corrections become significant beyond $z\approx 6.5$.

We now compare the comoving star formation rate deduced for $z\approx
6$ galaxies based on our candidate $i'$-dropout source counts with
predictions based on a range of rest-frame UV luminosity
functions. For convenience we assume that there is no evolution over
the sampled redshift range, $5.6<z<6.5$, spanned by our $i'$-drop
selection (equivalent to a range between $0.8-1.0\,h_{70}^{-1}$\,Gyr
after the Big Bang). We take as a starting point the luminosity
function for the well-studied Lyman-break $U$-dropout population,
reported in Steidel et al.\ (1999), which has a characteristic rest-UV
luminosity $L^{*}_{3}=15\,h^{-2}_{70}\,M_{\odot}\,{\rm yr}^{-1}$ and a
characteristic comoving number density at $z\approx 3$ is
$\Phi^{*}_{3}=0.0014\,h^{3}_{70}\,{\rm Mpc}^{-3}\,{\rm mag}^{-1}$ in
our cosmology. The faint end slope of the Schechter function at
$z\approx 3$ is relatively steep ($\alpha=-1.6$) compared with
$\alpha=-1.0$ to $-1.3$ for lower-redshift galaxy samples (e.g.,
Blanton et al.\ 2003).

A tantalizing result from our work so far is that at $z\sim 6$ there
are far fewer UV-luminous star forming galaxies than would have been
predicted if there was no evolution, based on a comparison to the
well-studied $z\sim 3-4$ Lyman break population (Steidel et al.\
1999).  In fact, the volume averaged (comoving) star formation rate in
galaxies with $>15\,h^{-2}_{70}\,M_{\odot}\,{\rm yr}^{-1}$ is $\approx
6\times$ {\em less} at $z\approx 6$ than at $z\approx 3$. So it would
appear that at this crucial epoch, where abundant sources of UV
photons are needed to reionize the Universe, there is in fact a
deficit of luminous star forming galaxies.  We have conclusively shown
that the bright end of luminosity function has evolved greatly from
$z\sim 6$ to $z\sim 3$. Other groups have claimed less dramatic
evolution or even no evolution in the total volume-averaged star
formation rate, based on the same fields (Giavalisco et al.\ 2004;
Dickinson et al.\ 2004) and similar {\em HST}/ACS data sets (Bouwens
et al.\ 2003; Yan, Windhorst \& Cohen 2003), but these groups work
closer to the detection limit of the images and introduce large
completeness corrections for the faint source counts. To do a complete
inventory of the UV light from star formation we must address the
contribution of low-luminosity star-forming galaxies to the ionizing
flux.  The public availability of the Hubble Ultra Deep Field (UDF;
Beckwith, Somerville \& Stiavelli) can address this puzzle by pushing
down the luminosity function at $z\approx 6$ to well below the
equivalent of $L^{*}$ for the $z\approx 3$ population.

\section{The Hubble Ultra Deep Field}

The Hubble Ultra Deep Field (UDF) was a Cycle 12 STScI Director's
 Discretionary Time programme executed over September 2003 -- January
 200, comprising 400 orbits in 4 broad-band filters (including F775W
 $i'$-band for 144 orbits; F850LP $z'$-band for 144 orbits). The UDF
 lies within the Chandra Deep Field South (CDF-S).  As the UDF
 represents the deepest set of images yet taken, significantly deeper
 than the $I$-band exposures of the Hubble Deep Fields (Williams et
 al.\ 1996), and adds the longer-wavelength $z'$-band, it is uniquely
 suited to the goals of our program.

We performed the first analysis of $i'$-drops in the UDF (Bunker et
al.\ 2004), presenting details of 54 candidate star forming galaxies
at $z\approx 6$ in a preprint the day after the release of the UDF
data. Our analysis was subsequently independently repeated by Yan \&
Windhorst (2004), with agreement at the 98\% level (see Bunker \&
Stanway 2004 for a comparison). We take our magnitude limit as
$z'_{AB}<28.5$ (a $10\,\sigma$ cut).We measure a star formation
density of $0.005\,h_{70}\,M_{\odot}\,{\rm yr}^{-1}\,{\rm Mpc}^{-3}$
at $z\approx 6$ from galaxies in the UDF with
SFRs\,$>1.5\,h_{70}^{-2}\,M_{\odot}\,{\rm yr}^{-1}$ (equivalent to
$0.1\,L^{*}_{UV}$ at $z\approx 3$).

At the relatively bright cut of $z'_{AB}<25.6$ used in Stanway, Bunker
\& McMahon (2003) from the GOODSv0.5 survey, the UDF data is 98\%
complete for sources as extended as
$r_{h}=0.5$\,arcsec. Interestingly, we detect no extended (low surface
brightness) $i'$-drops to this magnitude limit in addition to the
targetted $i'$-drop SBM03\#1 in the deeper UDF data. This supports our
assertion that the $i'$-drop population is predominantly compact and
there cannot be a large completeness correction arising from extended
objects (c.f.\ Lanzetta et al. 2002). The ACS imaging is of course
picking out H{\scriptsize II} star forming regions, and these
UV-bright knots of star formation are typically compact ($<1$\,kpc,
$<0.2$\,arcsec at $z\approx 6$) even within large galaxies at low
redshift.

It is interesting that the level of stellar contamination in the UDF
$i'$-drops is only 2\%, compared with about one in three at the bright
end ($z'_{AB}<25.6$; Stanway, Bunker \& McMahon 2003; Stanway et al.\
2004). This may be because we are seeing through the Galactic disk at
these faint limiting magnitudes to a regime where there are no stars
(see also Pirzkal et al.\ 2005).

\section{Implications for Reionization}

We compare our $i'$-drop luminosity function with the work of Madau,
Haardt \& Rees (1999) for the density of star formation.  We have
updated their equation 27 for the more recent concordance cosmology
estimate of the baryon density of Spergel et al.\ (2003), and for the
predicted mean redshift of our sample ($z=6.0$):
\begin{equation}
{\dot{\rho}}_{\rm SFR}\approx \frac{0.026\,M_{\odot}\,{\rm yr}^{-1}\,{\rm Mpc}^{-3}}{f_{\rm esc}}\,\left( \frac{1+z}{7}\right) ^{3}\,\left( \frac{\Omega_{b}\,h^2_{70}}{0.0457}\right) ^{2}\,\left( \frac{C}{30}\right)
\end{equation}

This relation is based on the same Salpeter Initial Mass Function as
we have used in deriving our volume-averaged star formation rate.  $C$
is the concentration factor of neutral hydrogen, $C=\left<
\rho^{2}_{\rm HI}\right> \left< \rho_{\rm HI}\right>
^{-2}$. Simulations suggest $C\approx 30$ (Gnedin \& Ostriker 1997).
The escape fraction of ionizing photons ($f_{\rm esc}$) for
high-redshift galaxies is highly uncertain (e.g., Steidel, Pettini \&
Adelberger 2001), but even if we take $f_{\rm esc}=1$ (no absorption
by H{\scriptsize~I}) this estimate of the star formation density
required is a factor of $\approx 5$ higher than our measured star
formation density of $0.005\,h_{70}\,M_{\odot}\,{\rm yr}^{-1}\,{\rm
Mpc}^{-3}$ at $z\approx 6$ from galaxies in the UDF with
SFRs\,$>1.5\,h_{70}^{-2}\,M_{\odot}\,{\rm yr}^{-1}$.  For faint end
slopes of $\alpha ~-1.8\rightarrow-1.3$ galaxies with $L>0.1\,L^{*}$
account for $32-80$\% of the total luminosity, so would fall short of
the required density of Lyman continuum photons required to reionize
the Universe.  If the faint-end slope is as steep as $\alpha\approx
-1.9$ then there would just be enough UV Lyman continuum photons
generated in star forming galaxies at $z\approx 6$ (assuming a
Salpeter IMF), but the required escape fraction for complete
reionization would still have to be implausibly high ($f_{esc}\approx
1$, whereas all high-$z$ measurements to date indicate $f_{esc}\ll
0.5$: Fern\'{a}nadez-Soto, Lanzetta \& Chen 2003; Steidel, Adelberger
\& Pettini 2001).

AGN are also under-abundant at these epochs (e.g., Dijkestra, Haiman \&
Loeb 2004). If star forming galaxies at redshifts close to $z=6$ were
responsible for the bulk of reionization, then a very different initial
mass function would be required, or the calculations of the clumping
factor of neutral gas would have to be significantly over-estimated
(see also Stiavelli, Fall \& Panagia 2004).
Alternatively another low-luminosity population (e.g., forming globular
clusters; Ricotti 2002) could be invoked to provide some of the
shortfall in ionizing photons.  It is also plausible that the bulk of
reionization occurred at redshifts well beyond $z=6$: the WMAP
polarization data indicate  $z_{reion}>10$ (Kogut et al.\ 2003), and it
is possible that the Gunn-Peterson troughs seen in $z> 6.2$ QSOs (Becker
et al.\ 2001; Fan et al.\ 2002) mark the very last period of a neutral
IGM.

We have recently discovered evidence of an early epoch of star
formation (Eyles et al.\ 2005). Some of the
spectroscopically-confirmed $i'$-drops at $z\approx 6$ show evidence
of another spectral break in the near-infrared (comparing {\em
Spitzer} images with IRAC at $3.6-4.5\,\mu$m with {\em HST} NICMOS \&
ACS images at $0.8-1.6\,\mu$m). This age-sensitive Balmer/4000\,\AA\
break indicates the IR light is dominated by stars older than 100\,Myr
(and most probably 400\,Myr old). This pushes the formation epoch for
these galaxies to $z_{\rm form}=7.5-13.5$. Our recent results also
indicate that there were at least some galaxies already assembled with
stellar masses $\approx 3\times 10^{10}\,M_{\odot}$ (equivalent to
$0.2\,M^{*}$ today) within the first billion years. The early
formation of such systems may have played a key role in reionizing the
Universe at $z\sim 10$.

\begin{figure}
\resizebox{0.48\textwidth}{!}{\includegraphics{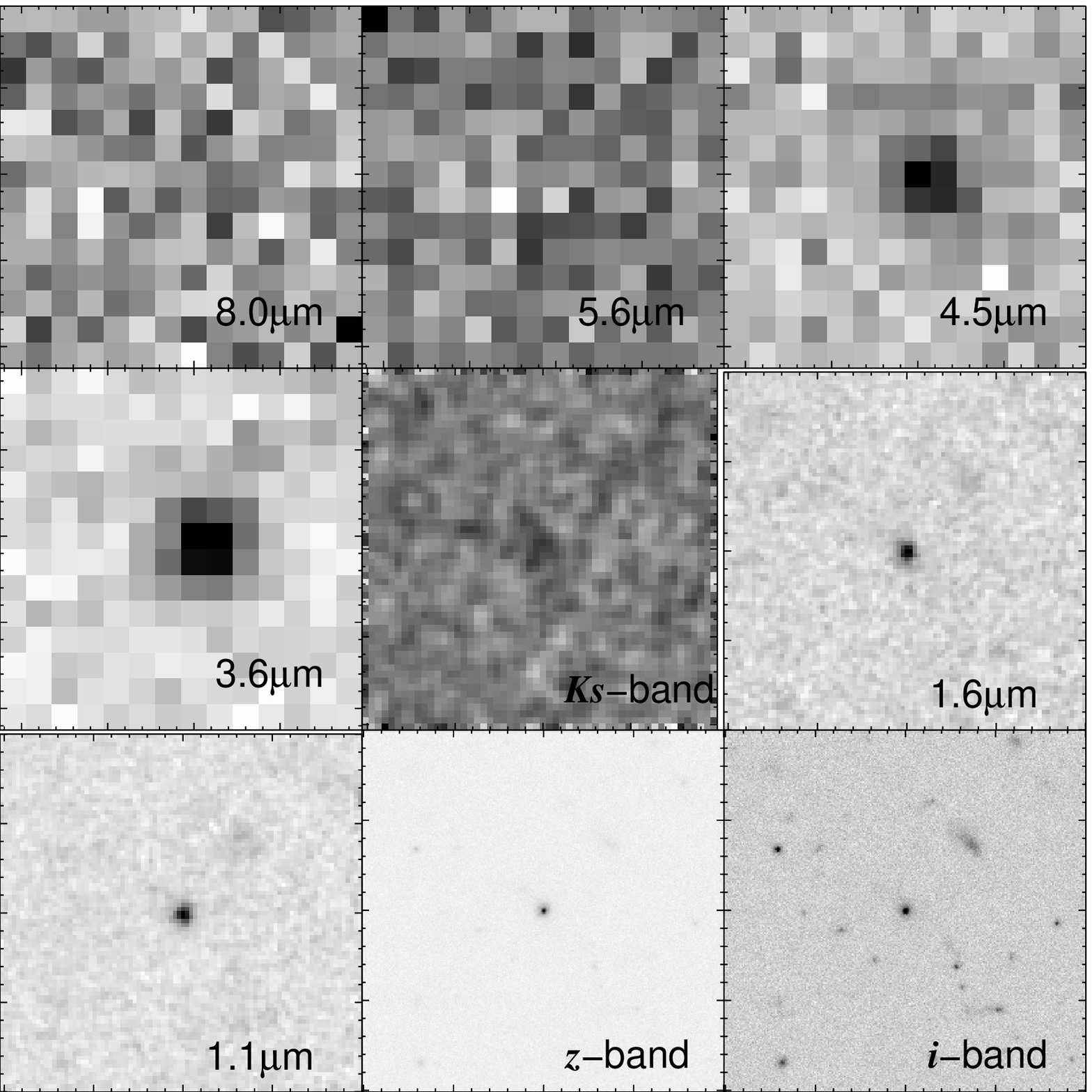}}
\resizebox{0.50\textwidth}{!}{\includegraphics{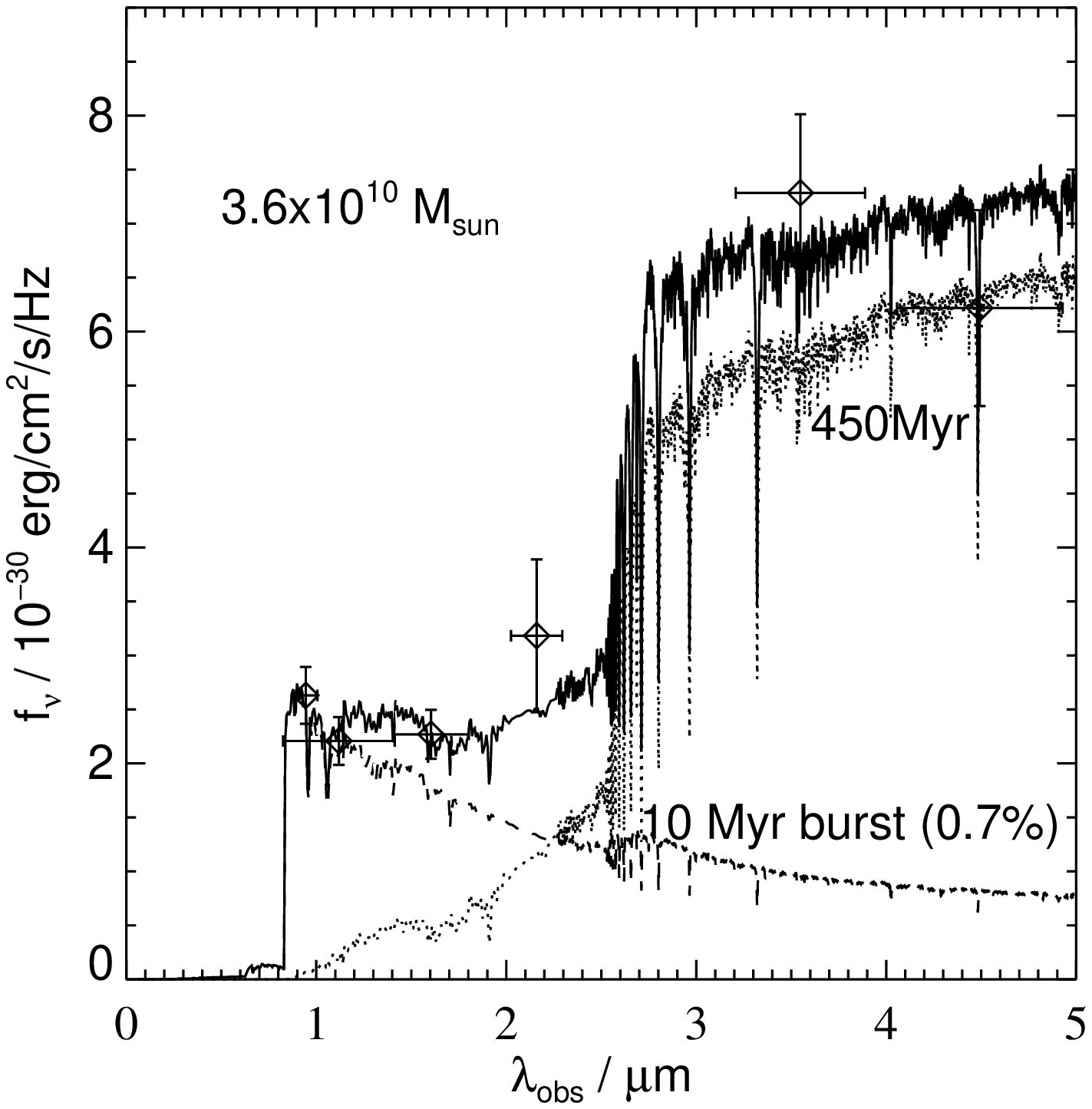}}
\caption{{\bf Left:} The {\em Spitzer} IRAC and {\em HST} ACS \&
NICMOS images of the $z=5.83$ galaxy, SBM03\#1. {\bf Right:} the {\em
Spitzer} and {\em HST} wavebands straddle the age-sensitive
Balmer/4000\,\AA\ break, and reveal an underlying old ($\sim
400$\,Myr) population which dominates the stellar mass, $\sim 3\times
10^{10}\,M_{\odot}$ (Eyles et al.\ 2005).}
\label{fig:spitzer}
\end{figure}

\end{document}